# REMOTE MONITORING OF WEAK AFTERSHOCK ACTIVITY WITH WAVEFORM CROSS CORRELATION: THE CASE OF THE DPRK SEPTEMBER 9, 2016 UNDERGROUND TEST

Bobrov D.I., I.O. Kitov, and M.V. Rozhkov

**Abstract**

The method of waveform cross correlation (WCC) allows remote monitoring of weak seismic activity induced by underground tests. This type of monitoring is considered as a principal task of on-site inspection under the Comprehensive nuclear-test-ban treaty. On September 11, 2016, a seismic event with body wave magnitude 2.1 was found in automatic processing near the epicenter of the underground explosion conducted by the DPRK on September 9, 2016. This event occurred approximately two days after the test. Using the WCC method, two array stations of the International Monitoring System (IMS), USRK and KSRS, detected $P_n$-wave arrivals, which were associated with a unique event. Standard automatic processing at the International Data Centre (IDC) did not create an event hypothesis, but in the following interactive processing based on WCC detections, an IDC analyst was able to create a two-station event with local magnitude ML=1.8. Location and other characteristics of this small seismic source indicate that it is likely an aftershock of the preceding explosion. Building on the success of automatic detection and phase association, we carried out an extended analysis, which included later phases and closest non-IMS stations. The final cross correlation solution uses four stations, including MDJ (China) and SEHB (Republic of Korea), with the epicenter approximately 2 km to north-west from the epicenter of the Sept. 9 test. We also located the aftershock epicenter by standard IDC program LocSAT using the arrival times obtained by cross correlation. The distance between the DPRK and LocSAT aftershock epicenters is 25.5 km, i.e. by an order of magnitude larger than that obtained by the WCC relative location method.

## Introduction

Seismic network of the International monitoring system (IMS) is monitoring the compliance with the Comprehensive nuclear-test-ban-treaty (CTBT). The IMS uses modern methods of data recording, acquisition, and transfer. The data are collected by the International data centre where they are processed with standard techniques. New advanced methods of data analysis, such as the method of waveform cross correlation (WCC), allow significant enhancement of monitoring capabilities by reducing the magnitude threshold of nuclear test detection at regional and teleseismic distances. Moreover, remote detection and location of weak aftershock activity becomes feasible.

In addition to significant reduction in detection threshold, the method of waveform cross correlation is characterized by dramatic improvement in the accuracy of relative location as well as magnitude estimation of small events compared to standard seismological methods [1]. In this paper, we present an example of superior WCC performance. It allowed to automatically detecting a low-magnitude seismic event that occurred on 11.09.2016 at 01:50:49.83 (UTC) near the epicenter (41.299°N, 124.049°E) of underground test conducted by North Korea on September 9, 2016 at 00:30:00.87, as estimated by the International Data Centre (IDC). At the same time, routine automatic processing did not create an event hypothesis, and thus, missed this event. Location and other characteristics of this small seismic source indicate that it is likely an aftershock of the preceding explosion. Therefore, the WCC provides unique and crucial information on the post-seismic processes induced by the announced underground test.

According to data from two seismic arrays of the International Monitoring System (IMS) of CTBT Organization (CTBTO), USRK (Russia) and KSRS (Republic of Korea), body-wave magnitude of the found event is 2.1. Location, seismic energy and the time elapsed since the moment of testing suggest the possibility of a causal link with the explosion. Therefore, it is possible to consider this event as an aftershock of the explosion. Magnitude 2.1 corresponds to an underground explosion of no more than ten to twenty tons of TNT, which determines the threshold for the biggest clandestine tests within the DPRK test site. Almost any event with magnitude of 2 or higher can be detected by the nearest IMS stations using cross correlation with signals from several previous explosions. Seismic waves from five announced DPRK underground tests were measured by IMS stations [1] and many other seismic networks - from global [2] to local [3]. According to the IDC, body-wave magnitude, $m_b$(IDC), of the smallest of the five tests was 4.1, and the biggest test was conducted on September 9, 2016 and reached magnitude 5.1.

Aftershocks associated with collapsing roof of the explosion cavity and induced release of tectonic energy are often observed as mechanical effects of underground explosions. For an explosion of magnitude ~ 5, that approximately corresponds to ten kilotons of TNT fired in hard rocks, induced seismic events usually have magnitude of two to three units lower than the magnitude of the triggering event [4,5]. Based on our experience with measurements after the explosions in different depths of burial and geological-geophysical conditions we can assume that the largest magnitude of aftershocks induced by five tests conducted in North Korea is unlikely to exceed 2-2.5 units. Even the best array stations with low ambient microseismic noise are unable to record signals from such events at teleseismic distances. At distances of several hundred kilometers, detection of signals from low-energy sources is possible, although it depends on seismic velocity and attenuation characteristics of the crust and upper mantle, as well as on microseismic noise variations. In the most favorable conditions, some regular regional phases can be observed even at three-component (3-C) stations.

In this study, we compare the results of signal detection, phase association as well as relative location and magnitude estimation obtained by the WCC method and in routine IDC processing. Both techniques are applied to the signals generated by the low-magnitude aftershock and measured by IMS arrays and non-IMS three-component stations. Qualitatively, the difference can be expressed in terms of finding this event by the WCC method and failure to find it in routine IDC processing. Quantitatively, the number of associated detections for the studied event, signal-to-noise ratios (SNR) of the detected signals, and location errors best illustrate the advantage of the WCC method.

**Data and method**

Since October 9, 2006, the date of the first DPRK test, standard detection methods used by the IDC we unable to find any seismic event that could be associated with the area of the DPRK explosions, except the explosions themselves. To increase the resolution of the IMS network data, since 2011 the IDC specialists have been testing the performance of the waveform cross correlation method using continuous comparison of current data with template waveforms from the previous DPRK explosions [1,6]. Cross correlation coefficient, *CC*, is used as a quantitative measure of the closeness between two signals. It is calculated as cosine of the angle between two vectors - the template record and current segment of the same length. For an array station, individual cross correlation coefficients are first calculated for each element $j = 1,…, M$, where *M* is the number of elements of the array. Then we average individual $CC_j$ and obtain the aggregate *CC* value:

$$CC = \sum_{j=1}^{M} CCj(t)/M$$

Averaging of the coherent *CC*-traces improves the signal-to-noise ratio more effectively than standard beam forming [1]. Another tool to improve SNR is waveform filtering. Before we calculate *CC*-traces, all waveforms from individual channels of arrays and 3-C stations are filtered using a set of four causal band-pass filters: 0.8 to 2 Hz, 1.5 to 3.0 Hz, 2.0 to 4.0 Hz, and 3.0 to 6.0 Hz. For automatic processing, the length of cross correlation windows are defined by frequency bands: 4.5 s for the highest frequency band and 6.5 s for other bands. In interactive analysis, we vary the window length to obtain optimal detection and to calculate signal attributes.

On September 11, 2016 the WCC-based detector found $P_n$-wave arrivals at two IMS array stations, USRK and KSRS, which are located at distances of ~ 410 km and ~ 440 km from the epicenter of the fifth DPRK explosion (DPRK5), respectively. Fig. 1 shows seismograms, which include signals from the DPRK5 and the aftershock. The most convincing evidence of the spatial closeness of the aftershock and explosion consists in almost complete coincidence in shape and relative arrival times of different phases at channels BH1 (E-W) and BHZ at USRK (Fig. 1c). The $L_g$-waves at station KSRS are also similar at BHE (Fig. 1a). Signal amplitudes from two sources differ by a factor of 550 or by 2.7 magnitude units. The amplitudes of signals from the aftershock at these two stations are so small that the event hypothesis cannot be created in IDC automatic processing because the estimate of magnitude of such an event would be below the preset threshold.

Several years ago we developed a cross correlation based procedure for signal detection and creation of event hypotheses (i.e. phase association) [1], which is currently used in tentative processing, with the explosion conducted on Jan 6, 2016 (DPRK4) serving as a master-event. As in routine IDC processing, the WCC technique is using the STA/LTA detector, which is already implemented at the IDC for original waveforms. This detector is based on a running short-term-average (STA) and long-term-average (LTA) computed recursively using previously computed STA values. The LTA lags behind the STA by a half of the STA window. The length of the STA and LTA windows have to be defined empirically as associated with spectral properties of seismic noise and expected signal. We have carried out a thorough investigation and determined the following windows: 0.8 s for the STA and 20 s for the LTA. The detection threshold is set to 2.5 [1]. The STA/LTA ratio is also considered as the signal-to-noise ratio (SNR). Since the STA/LTA is estimated from the *CC*-traces instead of the original waveforms the corresponding signal-to-noise ratio is called $SNR_{CC}$.

a)

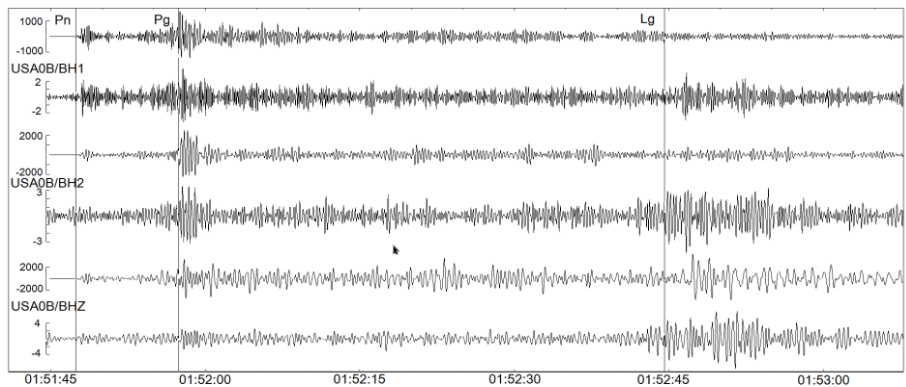

b)

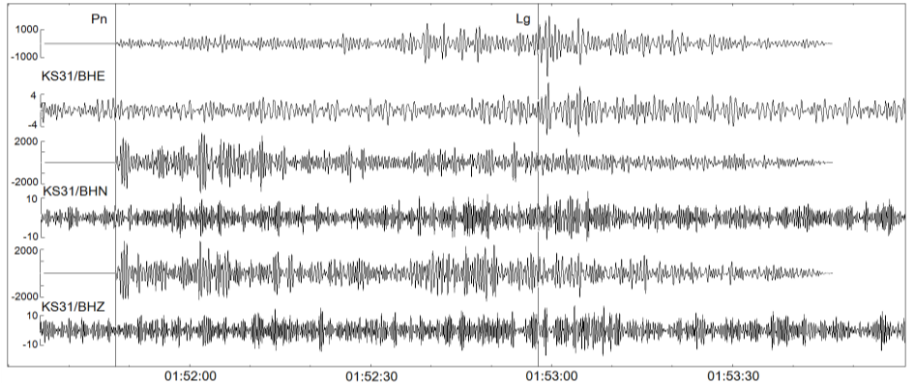

c)

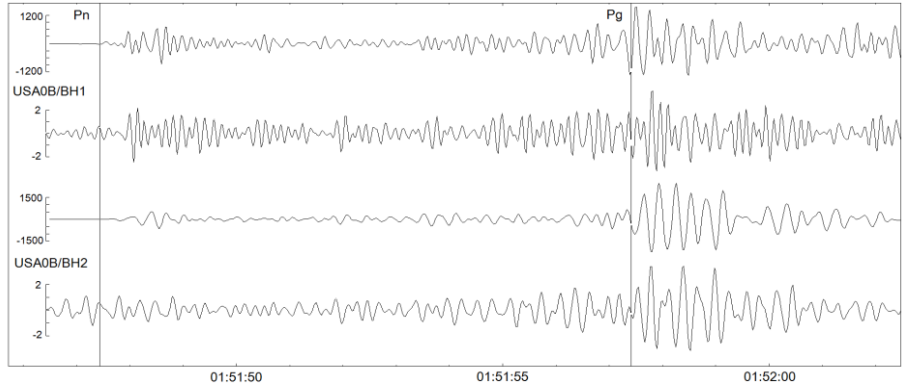

Fig. 1. Component-to-component, E-W (H1), N-S (H2), Z, comparison of signals from the DPRK5 and its aftershock as measured at IMS stations KSRS (a) and USRK (b). $P_n$- and $P_g$-waves from the aftershock at channels H1 (E-W) and H2 (N-S) of USRK are similar to those from the DPRK5 - c).

Cross correlation coefficient, especially for weak signals, degrades rapidly with distance increasing from the master. Correspondingly, the quality of detections falls and the rate of false alarms is growing. There is a technique reducing the false alarm rate. At array stations, we use the *CC*-traces at individual sensors and F-K analysis to calculate azimuth and slowness for all detected signals. Only the arrivals with azimuth and slowness residuals less than $\pm 20°$ and $\pm 2$ s/deg, respectively, are used for phase association. This procedure removes most of false detections. When all inappropriate arrivals are screened out, we have a set of detections with their absolute arrival times for each station, $t_{ij}$, where *i* is the index of the *i*-th arrival at station *j*. Considering the aftershock as a very small event with weak signals, one may assume that two detected $P_n$-wave arrivals could only be associated with a seismic event in the immediate vicinity of the DPRK5.

Initial stage of phase association procedure consists in reduction of arrival times to origin times. If an event is spatially close to the master, the travel times to all relevant stations can be predicted from the empirical master/station travel times, $tt_j$. For example, the origin times of potential slave events within the DPRK test site are calculated at several IMS stations as the difference between arrival times and the travel times for a given seismic phase from the DPRK tests [1,7]. Thus, by using empirical travel times one can calculate the approximate origin times, $ot_{ji}$, for all detections:

$$ot_{ji} = t_{ij} - tt_j$$

As a result, the set of arrival times is reduced to a set of origin times. When two or more origin times create a tight group, they can be associated with a source. When one or several origin times can be associated with different sources, it is necessary to build and test various hypothesis. Finally, the final set of built events should associate each arrival with not more than one event. The event origin time is the averaged origin times of all associated arrivals. Since the WCC-method is applicable to slave events, which are local to the master-event, and thus, uses only empirical master-station travel times for phase association we call this process *Local Association* (LA).

With the increasing master/slave distance (say, beyond 10 to 20 km), the master-station travel time loses its predictive power. One has to correct the travel times for the slave in the above equation. This improves the process of origin times' association. For slave event potential locations, we have introduced a mesh around the master event in a form of five circles spaced by 15 km. By virtue of design, the distance between circles becomes the accuracy of the slave location in WCC automatic processing. The first circle has 6 nodes, where the travel times are corrected for the slave/master distance, and the number of nodes doubles, triples, etc. for the next circles. In each node, the arrival times are reduced to origin times using the theoretical travel times correction corresponding to the relative node location. When arrival times are accurately estimated, the search over all nodes should give the smallest RMS origin time residual for the node likely closest to the sought slave event. Figure 2 depicts a grid for the Local Association with the DPRK4 in the center. Grid colors correspond to the RMS differences between the predicted and observed origin times. The automatically built two-station event hypothesis belongs to the node shown by red star, which is 15 km far from the DPRK4. The RMS residual in this point is 0.054 s. This hypothesis is not in the center of the grid because the arrival times were determined in an automatic process with lower resolution and the grid is rather sparse. Interactive WCC analysis can improve location as discussed in the following section.

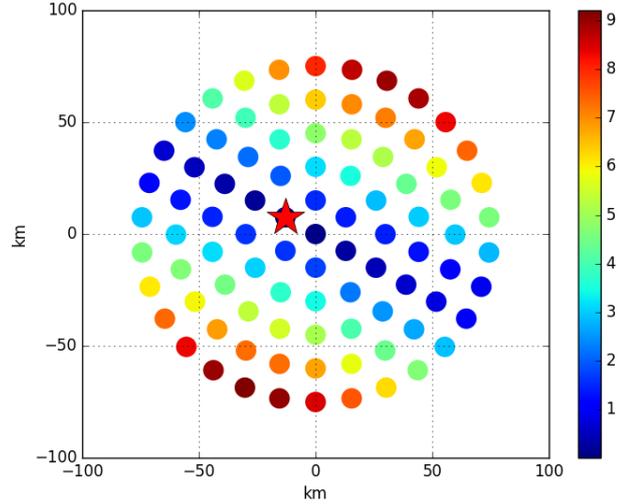

Fig. 2. The local association grid around the DPRK4. The distance between circles and nodes is approximately 15 km. Red star shows the node with the automatically found event hypothesis. The RMS origin time residual for the two-station event is 0.054 s.

Table 1 lists station data from the automatic bulletin for the found event, including arrival times, cross correlation coefficients, signal-to-noise ratios ($SNR_{CC}$), the travel time residuals, **$t_{res}$**, and the estimates of relative event magnitude, *dRM*. The latter is based on the ratio of signal norms: $|\mathbf{x}|/|\mathbf{y}|$, where **x** and **y** are the vector data (with the CC-window length) of the slave and master, respectively. The logarithm of the ratio,

$$dRM = \log(|\mathbf{x}|/|\mathbf{y}|) = \log|\mathbf{x}| - \log|\mathbf{y}|,$$

is the magnitude difference between two events or *relative magnitude*. This difference has a clear physical meaning for close events with similar waveforms. For a given slave event, the relative magnitude is a reliable dynamic parameter for a correct arrival association at several stations. Phase association is possible if the deviation of station relative magnitude at a given station from the network-averaged value does not exceed 0.7. As shown in Table 1, the average relative magnitude is *RM*=-2.69. For the estimated DPRK4 magnitude of $m_b$(IDC)=4.82 the body wave magnitude $m_b$(IDC) of the aftershock is 2.13.

After detecting signals from the aftershock at two stations, we carried out a thorough search of secondary phases and additional arrivals at different IMS stations. We reduced the detection threshold and extended the ranges allowed for the residuals of azimuth, slowness, and relative magnitude [1,8]. At USRK, we have found two later arrivals $P_g$-wave and $L_g$-wave, and at KSRS – only $L_g$-wave. At other IMS stations, signals from the aftershock were not detected by standard methods as well as by the WCC. The most likely candidates were array stations MJAR (Japan) and SONM (Mongolia) at distances of 960 km and 1930 km, respectively. Since other IMS stations are at teleseismic distances from the aftershocks, the probability of detection of events with magnitude of 2.1 was low.

Table 1. Automatic cross correlation bulletin for the DPRK5 aftershock with DPRK-4 as master-event

| Sta | Dist, km | EvStaAz, deg | Phase | Arrival time | $t_{res}$, s | CC | dRM | $SNR_{CC}$ |
|---|---|---|---|---|---|---|---|---|
| **USRK** | 410 | 35.8 | $P_n$ | 01:51:46.46 | 0.1 | 0.30 | -2.61 | 3.9 |
| **KSRS** | 440 | 193.6 | $P_n$ | 01:51:52.16 | -0.1 | 0.21 | -2.78 | 4.3 |

At regional distances, there are several three-component seismic stations not belonging to the IMS network with data available *via* IRIS (http://www.ds.iris.edu). All relevant IDC processing procedures were applied to these stations in order to detect signals from the aftershock. Two signals have been found at stations - SEHB (RK) and MJD (China), located at distances of 346 km and 367 km, respectively. Fig. 3 shows a map of the relative positions of the source area and four stations, and Fig. 4 compares the signals from the DPRK5 and its aftershock on SEHB and MJD. As for two IMS stations, secondary phases are most prominent on horizontal channels, but the $P_n$-wave arrival is not very clear. Detection of such signals is a challenge to the WCC method.

**Relative location**

The closeness of epicenters makes it possible to use the difference in travel times at several stations for accurate estimation of the relative positions of the DPRK5 and its aftershock [9]. Essentially, this the same minimization procedure of the dispersion of origin times as described in the previous section for local association but with a very small spacing between nodes. It is important to stress that the aftershock relative location is the estimation of the distance from the IDC absolute location of the DPRK5, but not the DPRK4 used in automatic processing. The result of relative location does not change the absolute location of the DPRK5, but the accuracy of the DPRK location is inherited by the aftershock. This method has to reduce the uncertainty of the aftershock absolute location by one-two orders of magnitude when compared with standard methods [10].

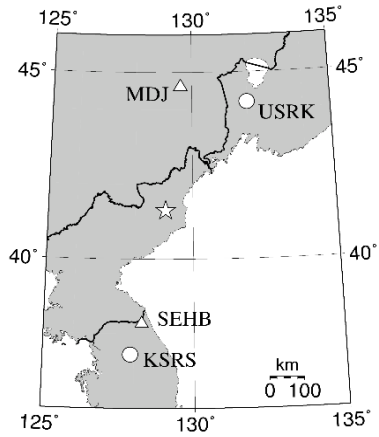

Fig. 3. Relative positions of the test site (star), two IMS arrays (open circles), and two 3-C stations (open triangles).

Stations USRK and MDJ are in almost opposite direction from the source area compared to KSRS and SEHB. Therefore, the aftershock epicenter shifting to the north with respect to the point of explosion will lead to a simultaneous increase in travel time to two southern stations and decrease in travel time to USRK and MDJ. When moving the aftershock epicenter in the east-west direction one cannot change much the difference in travel times. Hence, the resolution of the relative location is higher in the north-south direction.

For the accuracy of relative location, the accuracy of arrival time picking is particularly important. We have increased the resolution of time picking with the WCC method by resampling all records up to 200 Hz from 20 Hz at KSRS and 40 Hz for other three stations. To improve the arrival time estimates and to ensure their consistency we have varied the frequency content of all signals using various band-pass filters and changed the length of correlation windows. At all stations, we have found $P_n$-wave detections demonstrating stable arrival times within a few hundredths of a second that corresponds to a few hundred meters mislocation.

a)

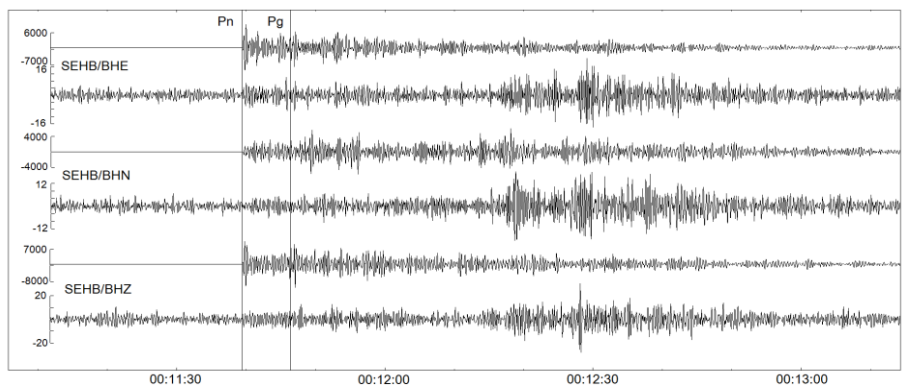

b)

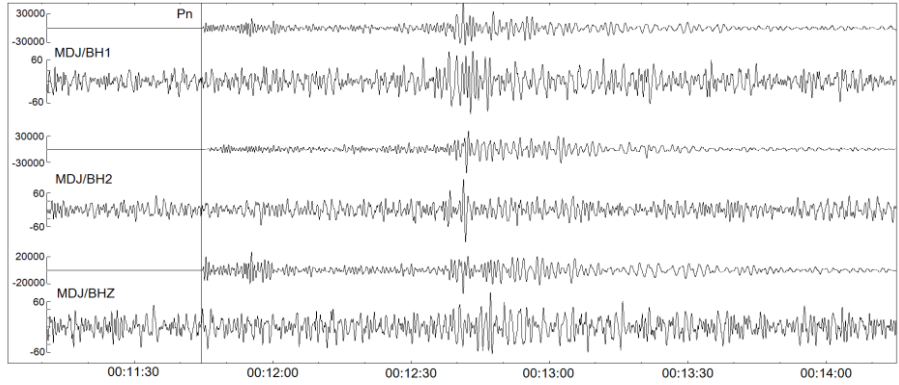

Fig. 4. Component-to-component, E-W (H1), N-S (H2), and Z, comparison of signals from the DPRK5 and its aftershock as measured at stations SEHB (a) and MDJ (b). The $P_n$-wave arrival at SEHB is poor, nevertheless found by WCC.

Table 2. Bulletin for the aftershock as obtained with the DPRK5 as a master event

| Sta | Dist, km | EvStaAz, deg | Phase | Arrival time | $t_{res}$, s | CC | dRM | $SNR_{CC}$ | SNR |
|---|---|---|---|---|---|---|---|---|---|
| **SEHB** | 346 | 191.7 | $P_n$ | 01:51:39.35 | -0.21 | 0.25 | -2.90 | 2.6 | 1.9 |
| **MDJ** | 367 | 6.7 | $P_n$ | 01:51:43.35 | 0.26 | 0.28 | -2.66 | 2.7 | 2.1 |
| **USRK** | 410 | 35.8 | $P_n$ | 01:51:46.05 | 0.16 | 0.26 | -2.94 | 3.1 | 2.8 |
| **KSRS** | 440 | 193.6 | $P_n$ | 01:51:51.92 | -0.21 | 0.20 | -2.89 | 3.3 | 2.2 |

The relative position of the aftershock is defined as the point minimizing the RMS travel time residual, **δt**, calculated as the difference between travel times from two sources to four stations, as described in the previous section. The only difference is that the node spacing in the relative location procedure is 50 m instead of 15 km used in the *LA*. Fig. 5 shows the distribution of **δt** for the DPRK5/aftershock pair. The aftershock epicenter is characterized by the minimum RMS residual (0.0023 s) and is located at a distance of approximately 2.1 km to the north-west from the DPRK5. Table 2 lists the arrival attributes from the event bulletin, which includes four stations defining the epicenter location. When the event depth is fixed to the surface, the origin time is estimated at 01:50:49.78. Limited geological information [11] indicates the possibility that the aftershock hypocenter is situated near the fault separating basalts and stratified volcanic rocks.

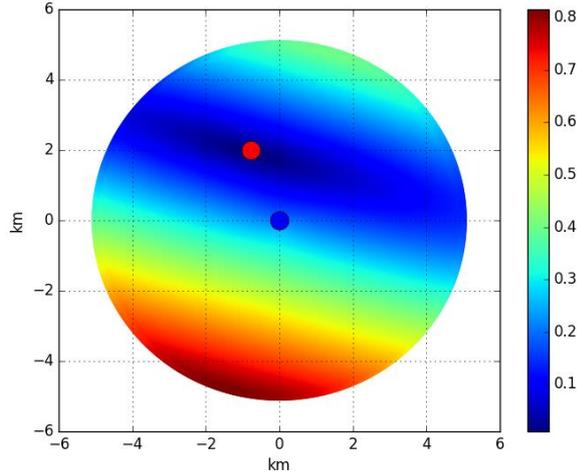

Fig. 5. The position of the aftershock (red dot) relative to the DPRK5 epicenter (blue dot). The distance between two events is ~2 km as estimated from cross correlation arrivals at four stations. Color bar shows the RMS origin time residual measured in seconds.

The average relative magnitude estimated by four stations in Table 2 is -2.84. For the DPRK5 magnitude of $m_b$(IDC)=5.09, absolute body wave magnitude of the aftershock is $m_b$(IDC)=2.25. The initial estimate from two stations in Table 1 is 2.13. The difference of 0.12 is likely related to the estimate of -2.66 at station SEHB. Figure 3 shows that the $P_n$-wave arrival at this station has practically the same amplitude as the ambient noise (SNR=1.9 in Table 2). Therefore, the relative magnitude estimate corresponds to noise rather than to signal, and thus, might be overestimated. Excluding the SEHB d*RM* estimate, we get the aftershock magnitude of 2.18, which is very close to the initial value.

Figure 6 illustrates the difference between the IDC location procedure and that using the WCC method. Two absolute locations obtained with IDC program LocSAT are shown: the DPRK5 (red star) and aftershock (blue circle), both are accompanied by 90% confidence ellipses. The relative location of the aftershock is shown by green circle. The distance between the DPRK and the LocSAT aftershock epicenters is 25.5 km, i.e. by an order of magnitude larger than that obtained by the WCC relative location method. The distance between two aftershock locations is 26.8 km. However, the relative aftershock location is within the LocSAT confidence ellipse so two solutions are consistent.

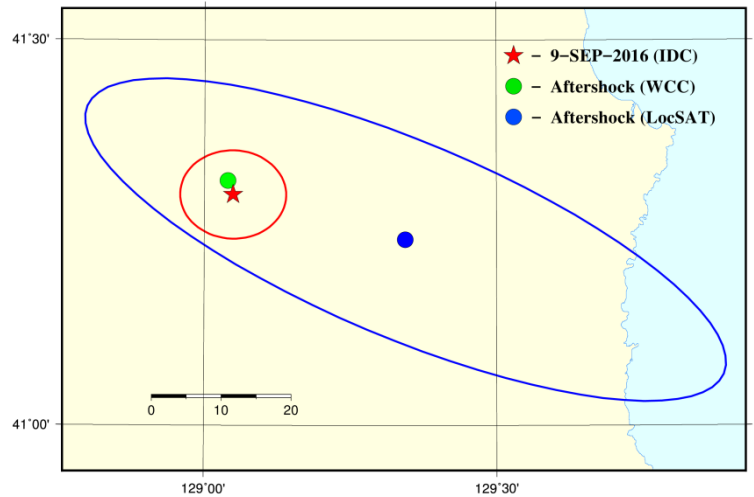

Fig. 6. Absolute locations of the DPRK5 (red star) and aftershock (blue circle) with 90% confidence ellipses as obtained by IDC location software LocSAT. The result of aftershock relative location is shown by green circle.

**Discussion**

The use of waveform cross-correlation at the International Data Centre of the CTBTO makes possible detecting seismic signals at arrays and three-component stations and their association with a magnitude 2 unique source within a few km from the epicenters of the DPRK5 explosion. Overall, we can interpret this small event as an aftershock, which would be missed if the tentative WCC detection and association procedure would not be used. Measuring abrupt stress relaxation processes after nuclear tests is crucial for the success of nuclear test monitoring under the CTBT. Standard IDC processing did not create an event hypothesis.

The improvement in detection due to the WCC method is best illustrated by the comparison of SNR values in the last two columns of Table 2. The estimates based on the WCC ($SNR_{CC}$) are consistently larger than those obtained by standard method. This effect is most important at 3-C stations where standard SNR estimates are lower. In essence, we can find smaller aftershocks of underground test or natural events using cross correlation with high-quality waveform templates from a number of master-events.

The improvement in accuracy of absolute location of small events is striking compared to that obtained by standard methods. For this particular case, the gain is approximately an order of magnitude. Moreover, the obtained distribution of the residual travel time in Figure 5 does not allow rejecting the hypothesis that the aftershock is closer to the epicenter of the explosion. Vast experience gained working with cross correlation at regional 3-C and array stations [1,10] indicates that the level of similarity between signals rapidly falls with distance from the master event. Therefore, the event we have found should be at a maximum distance of 2 to 5 km from the epicenter of the DPRK5.

The principal purpose of this study is to enhance the capability of the IMS seismic network serving for the CTBT monitoring regime. Theoretical improvements to the WCC technique, for example, new statistical methods describing the properties of signals and microseismic noise, should be accompanied by expansion of its practical use in various regional and global networks and by improvement in its resolution, e.g. by the use of three-component arrays [12]. One of the most important WCC elements consists in careful collection of data from all types of well-located historical seismic events, including analog and digital records from underground nuclear explosions.

**References**

[1] *Bobrov D., Kitov I., Zerbo L.* // Perspectives of cross-correlation in seismic monitoring at the International data centre. Pure and Applied Geophys. 2014. V. 171. N3-5. P.439-468

[2] *Barth, A*. // Significant release of shear energy of the North Korean nuclear test on February 12, 2013. Journal of Seismology. July 2014. V. 18, N 3. P. 605–615. doi:10.1007/s10950-014-9431-6

[3] *Adushkin V. V., I. O. Kitov, N. L. Konstantinovskaya, K. S. Nepeina, M. A. Nesterkina, and I.A. Sanina* // Detection of Ultraweak Signals on the Mikhnevo Small Aperture Seismic Array by Using Cross Correlation of Waveforms. Doklady Earth Sciences, 2015. V. 460. № 2. P. 189-191

[4] *Kitov I.O., O.P. Kuznetsov* // Reports of the Academy of Sciences USSR. 1990. V. 315. N4. P. 839-842. (in Russian)

[5] *Adushkin V.V., Spivak A.A.* / Geomechanics of large-scale explosions. M. Nedra. 1993. 319 p. (in Russian)

[6] *Kitov I., Rozhkov M., and Bobrov D.* // S31A-2703 *Proc. AGU Fall Meeting San-Francisco*, Dec. 12-16 Dec 2016

[7] *Bobrov D. I., I. O. Kitov, M. V. Rozhkov, and P. Friberg* // Seismic Instruments. 2016. V. 52, N 1, pp. 43–59.

[8] *Coyne J., Bobrov D., Bormann P., Duran E., Grenard P., Haralabus G., Kitov I., Starovoit Yu.* New Manual of Seismological Practice Observatory. 2012. Ch. 15. DOI: 10.2312/GFZ.NMSOP_2_ch15.

[9] *Bobrov D., Kitov I., and Rozhkov M.* // EGU2016-6620 Proceedings of the 2016 EGU General Assembly 2016. Vienna. April 17 – 22. 2016.

[10] *Richards P., F. Waldhauser, D. Schaff, and W.-Y. Kim* // Pure and Appl. Geophysics. 2006. V. 163. P. 351–372

[11] *Coblentz D., Pabian F.* // Science & Global Security: The Technical Basis for Arms Control, Disarmament, and Nonproliferation Initiatives. V.23 (2), 2015.

[12] *Adushkin V. V., I. O. Kitov, and I. A. Sanina* // Application of a Three Component Seismic Array to Improve the Detection Efficiency of Seismic Events by the Matched Filter Method. Doklady Earth Sciences, 2016. V. 466. N 1. P. 47-50